\documentclass[amsmath,preprintnumbers,10pt,twocolumn,prl]{revtex4-1}
\usepackage{amsbsy}
\usepackage{amsmath,amsfonts,bm}
\usepackage{graphicx}
\usepackage{color}
\usepackage{subfigure}

\usepackage{graphicx}

\graphicspath{{images_ReviewJSP/}}

\usepackage{verbatim}

\begin{document}
\title{Biological implications of dynamical phases in non-equilibrium  networks}
\author{Arvind Murugan${}^{1,2,3}$ and Suriyanarayanan Vaikuntanathan${}^{2,4}$}
\affiliation{${}^{1}$ School of Engineering and Applied Sciences and Kavli Institute for Bionano Science and Technology, Harvard University, Cambridge, MA 02138,\\${}^{2}$James Franck Institute, University of Chicago, ${}^{3}$Department of Physics, University of Chicago, \\ ${}^{4}$Department of Chemistry, University of Chicago, Chicago, IL 60637} 

\begin{abstract}
Biology achieves novel functions like error correction, ultra-sensitivity and accurate concentration measurement at the expense of free energy through Maxwell Demon-like mechanisms. The design principles and free energy trade-offs have been studied for a variety of such mechanisms. In this review, we emphasize a perspective based on dynamical phases that can explain commonalities shared by these mechanisms. Dynamical phases are defined by typical trajectories executed by non-equilibrium systems in the space of internal states. We find that coexistence of dynamical phases can have dramatic consequences for function vs free energy cost trade-offs. Dynamical phases can also provide an intuitive picture of the design principles behind such biological Maxwell Demons.
%Here, we review these mechanisms with an emphasis on 
\end{abstract}

\maketitle

\section{Introduction}
Cells constantly sense and process stimuli from the environment in order to adapt to changing conditions. Uncovering the mechanisms used by cells for information processing and organization is a crucial problem in systems biology and biophysics. The necessity to dissipate energy in order to achieve control or for information processing is now well understood\cite{Qian:2007gb}.

A classic example of a dissipative non-equilibrium mechanism in biology is kinetic proofreading\cite{Hopfield:1974uo,Hopfield:1980vm,Ninio:1975vv}. Kinetic proofreading allows crucial {biochemical} processes, such as protein synthesis, to proceed with error rates much lower than {those implied by equilibrium binding affinities}. Other prominent biological examples {of information processing} include sensory adaptation and ultra-sensitive switching - such as those responsible for controlling chemotaxis in E. Coli \cite{Tu:2013ws} - which modulate the response of the cell to external stimuli. Measurement of ligand concentrations by receptors, a problem first studied by Berg and Purcell, can also be enhanced by the consumption of free energy~\cite{Mehta:2012ji}.

These diverse mechanisms in biology share intriguing similarities. All of these mechanisms can be modeled as a non-equilibrium driven reaction networks using a master equation,
\begin{equation}
\partial_t p_i = \sum_j \mathbb{W}_{ji} p_j - \mathbb{W}_{ij} p_i
\label{eqn:masterEqn}
\end{equation} 
where $p_i$ is the occupancy of state $i$, $ \mathbb{W}$ is a transition matrix and the rates $ \mathbb{W}_{ij}$ break detailed balance; i.e., $\prod_{loop} \mathbb{W}_{ij}/\mathbb{W}_{ji} \neq 1$ around some loops in the network of states~\cite{Schnakenberg1976}. 

In each of the biological mechanisms listed above, we desire the sensitivity of the occupancy of a select state (or a set of states) $\mathcal{O}$ with respect to changes in the kinetic parameters $\mathbb{W}_{ij}$ to be greatly enhanced (or suppressed in the case of adaptation). Hence the biological function of these mechanisms can be described using a sensitivity $\nu$ of a set of states $\mathcal{O}$,
\begin{equation}
\nu \equiv \frac{\partial \log \sum_{i \in \mathcal{O}} p_i}{\partial \log X}
\label{eqn:nuGeneral}
\end{equation}

For example, in proofreading, $\log X$ represents a change in $
\mathbb{W}_{ij}$ due to a change in the energy of an enzyme-substrate bound state; in ultra-sensitivity \cite{Tu:2008vn}, adaptation\cite{Tu:2013ws} and concentration sensing \cite{Mehta:2012ji,Bialek:2005vq,Berg:1977uj}, $X$ represents changes in a ligand concentration \cite{Hartich:2015vp,Murugan:2014uq,Tu:2008vn,Lan:2012in}. In these processes, larger displacement from equilibrium (i.e., large detailed-balance breaking) typically leads to larger deviations of sensitivity $\nu$ from its equilibrium value~\cite{Lan:2012in,Tu:2008vn}.

\begin{figure}
\centering
\includegraphics[width=1\linewidth]{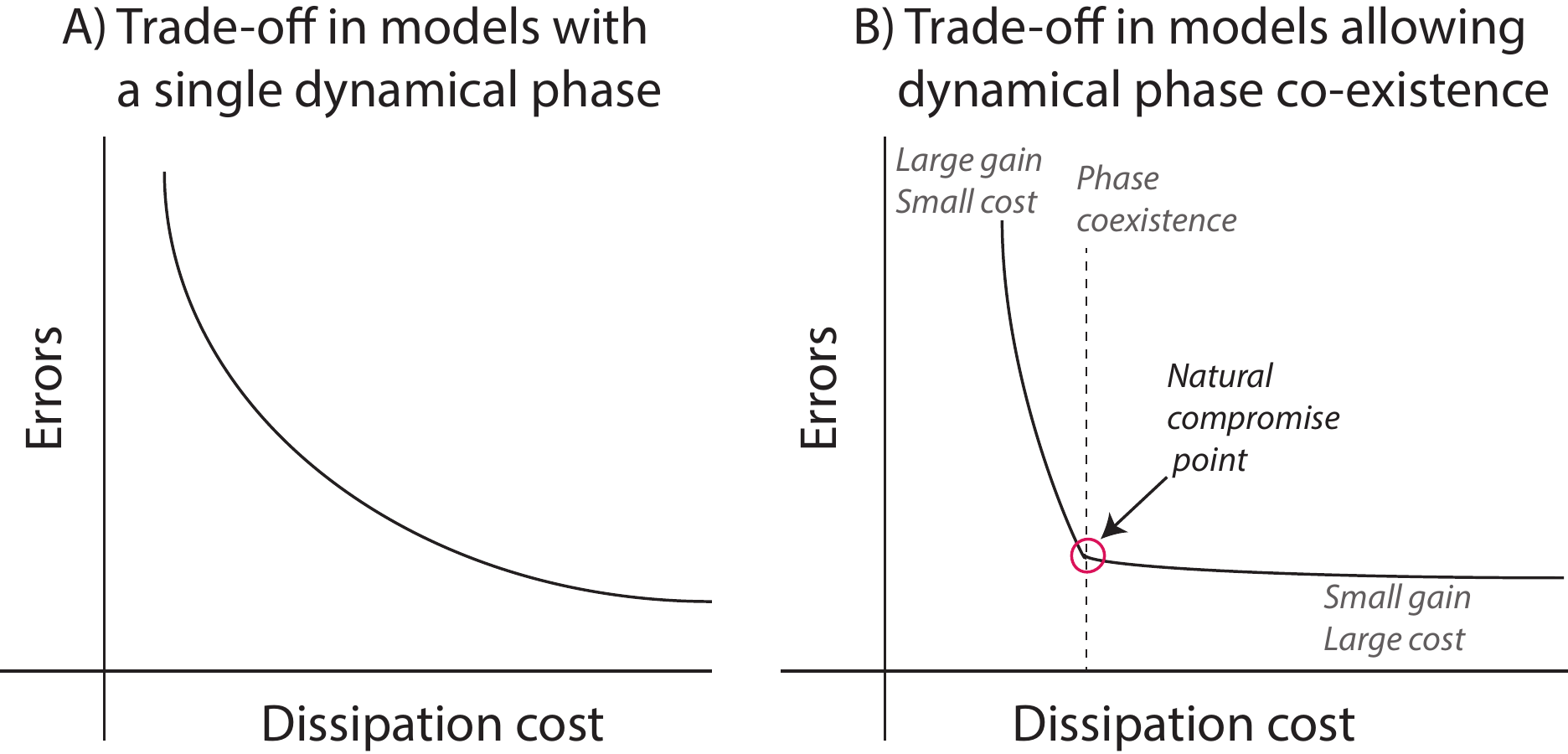}
\caption{Dynamical phase coexistence in non-equilibrium mechanisms can result in sharp turning points in the trade-off curve between function and dissipation. In kinetic proofreading (or error correction), coexistence separates a regime of dramatic gains in error correction at a small cost in dissipation from a regime of small gains at a large cost. Hence phase coexistence presents a natural compromise between the two incompatible goals of low errors and low dissipation. Similar considerations might apply to search strategies, accurate sensing of concentrations and other active processes.}
\label{fig:JSP_Tradeoffs}
\end{figure}

Thus, in all of these examples,  a biologically important function --- error correction, ultra-sensitive switching or sensory adaptation ---- is enhanced at the expense of free energy.  The trade-offs between energy dissipation and the enhancement of the particular relevant function have been derived for these different systems. Similarly, the detailed relationship between the kinetics of reaction networks and the enhancement of function have been studied for each of these systems.

Similarities in results across these systems raise natural questions: Are qualitative aspects of the trade-off between energy dissipation and enhancement of function intrinsic to the driven non-equilibrium nature of these systems?  Is there a minimal set of design principles that allow biological networks to have prescribed regimes of, say, ultra-sensitivity or proofreading ability? Can such design principles be formulated in terms of coarse-grained aspects of the network architecture? 
Such abstracted results are important both for synthetic design and to understand complex biochemical networks whose detailed kinetic constants might never be known. 
 
In this review, we emphasize a picture based on trajectories that might be useful in answering the above questions for diverse biological mechanisms. We show that the trajectories of non-equilibrium systems in state space can be classified as localized or de-localized. We discuss how being able to switch between such trajectories allows good function at significantly lower costs. We begin by first reviewing work on proofreading that shows all these themes; we then review how large deviation theory provides useful tools to study dynamical phases in such driven systems. Finally, we briefly discuss the relevance of these ideas to other systems like stochastic search problems, sensing of concentrations, ultra-sensitivity and sensory adaptation. 

\section{Dynamical phases and trade-offs in proofreading}

\begin{figure}
\centering
\includegraphics[width=0.8\linewidth]{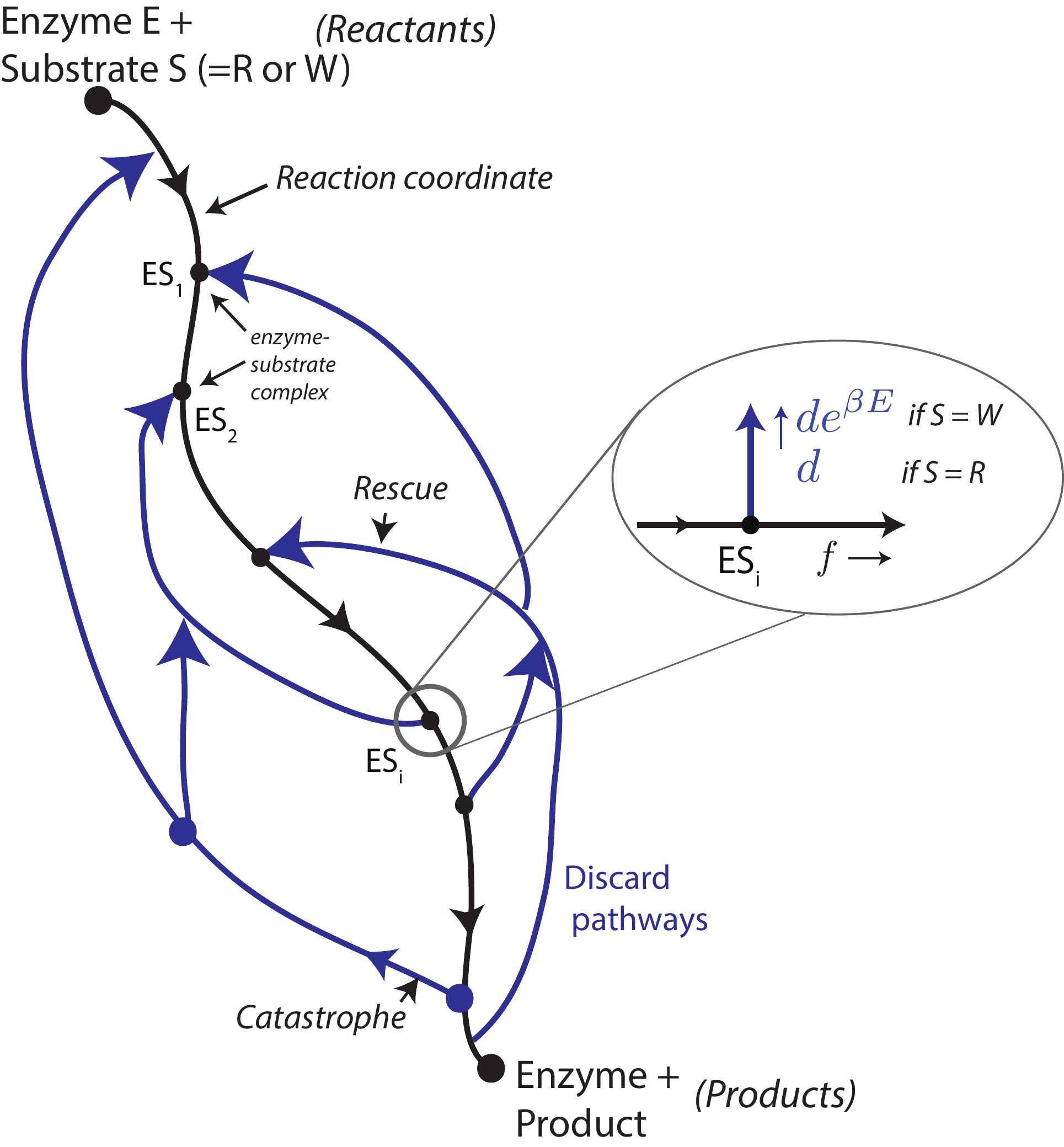}
\caption{Proofreading by a general network can be described using three ingredients. 1. A special pathway (reaction coordinate, black), exclusively through which the reaction is completed, 2. Catastrophes: the enzyme-substrate complex can take a discard pathway (blue) off the reaction coordinate and rapidly undo the progress made thus far 3. Rescues: instead of being reset completely to the reactants state, the system may be ``rescued'' to the reaction coordinate again, resulting in only a partial backtrack. Proofreading mechanisms turn small differences in the catastrophe (or rescue) rates of different substrates into large differences in the reaction completion rate. 
% (In more general contexts, the reaction coordinate (black) can represent a set of pathways through which the reaction is completed.)
}
\label{fig:JSP_discard_pathways}
\end{figure}

\begin{figure}
\centering
\includegraphics[width=0.7\linewidth]{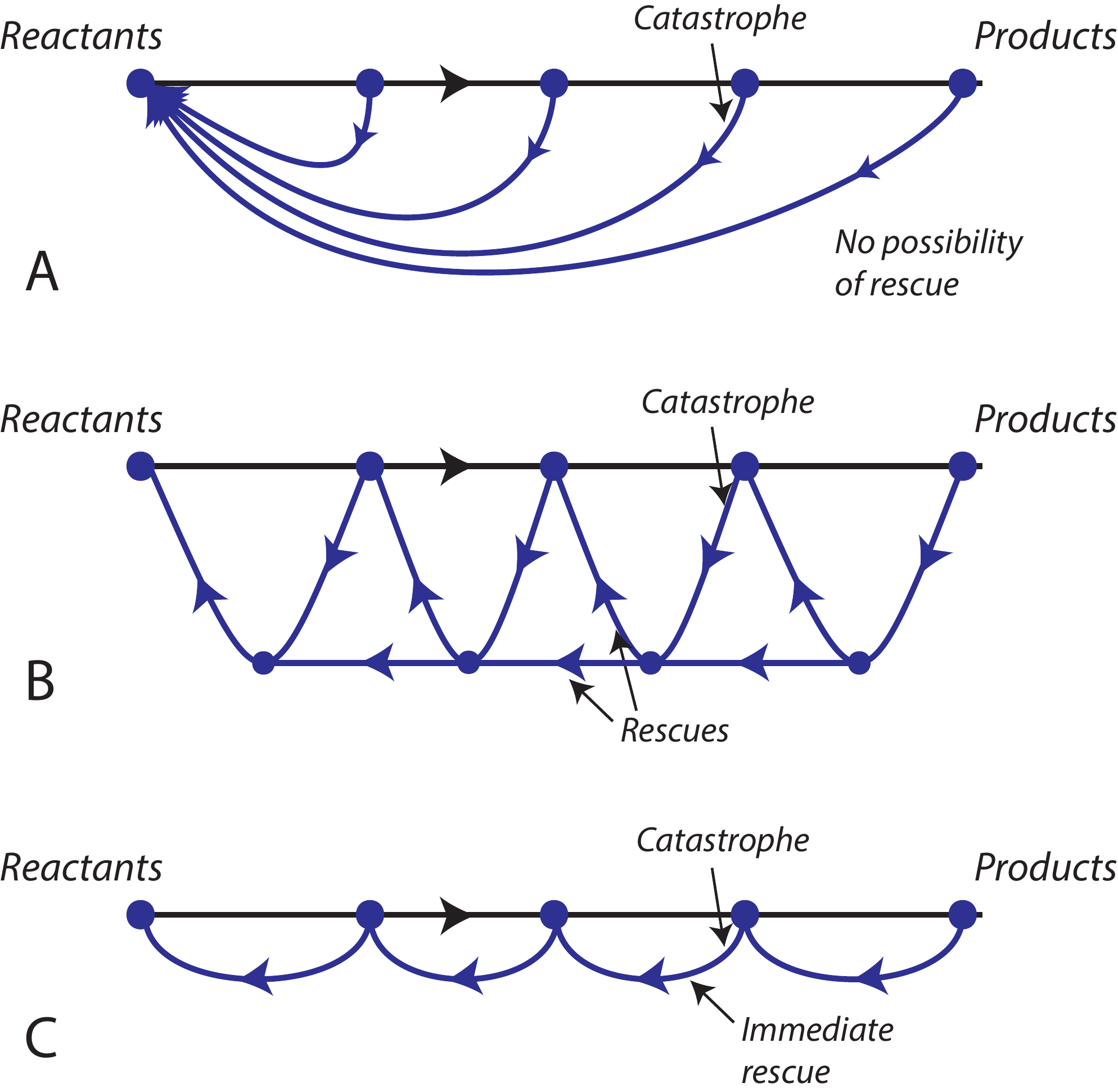}
\caption{Commonly studied proofreading networks with different rescue statistics. (A) Networks with no possibility of rescue are commonly used to model immune cell receptors, partly because of their simplicity. However, such networks can exhibit only one dynamical phase (localized at reactants) and hence have poor energy-error trade-offs (Fig.~\ref{fig:JSP_Tradeoffs}A) (B) Networks that support rescues throughout the system exhibit multiple dynamical phases. The rescue and catastrophe rates determine the distribution of backtracking lengths which in turn determines whether the system is de-localized, localized at reactants or localized at products. Such networks can exhibit the trade-off shown in Fig.~\ref{fig:JSP_Tradeoffs}B. (C) An extreme case of networks that favor rescues. The dynamical phases exhibited by this network are similar to those of (B); we use this network as a pedagogical example to illustrate calculations of dynamical phases. 
}
\label{fig:JSP_network_examples}
\end{figure}

%\textbf{Should I call them `rescue lengths' or `catastrophe lengths' or `backtracking length' .. = 
%backtracking length}

In the biological context, kinetic proofreading was one of the earliest examples of non-equilibrium dynamics proposed to improve an information processing function.  The original application of proofreading was to the problem of enzymatic specificity - how can an enzyme strongly discriminate between two structurally similar substrates? For example, an enzyme $E$ might be able process its native `right' substrate $R$ with kinetics $\mathbb{W}_{ij}^R$ (defined in Eqn.\ref{eqn:masterEqn}); however, $E$ might be able to process an undesirable but structurally similar (`wrong') substrate $W$ with kinetics $\mathbb{W}^W_{ij}$ that differ from $\mathbb{W}^R_{ij}$ by only a small extent. Can such a small difference in kinetics be magnified into a large difference in the net rate of the enzymatic reaction?

%For example, if an enzyme processes two similar substrates with slightly different kinetic constants $\mathbb{W}_{ij}$ , 

%For example, in the cell, the enzyme might be presented with `wrong' substrates $W$ that are structurally similar to the `right' substrate $R$; 

%We desire a large change in output (say, the occupancy of some state $N$) in response to such a small change in kinetics. 
For proofreading \cite{Sartori:2013fv}, this difference in kinetics is usually due to a difference in the enzyme-substrate binding energy. Hence we define a measure of discrimination between substrates of differing binding energy,
\begin{equation}
 \nu(E) \equiv - d \log p(E) / dE
 \label{eqn:nuE}
\end{equation}
where $p(E)$ is the occupancy of a final state whose energy $E$ is a measure of the enzyme-substrate binding energy and whose occupancy is representative of the rate of product formation \cite{Hopfield:1974uo}.

%We can formalize the problem of proofreading as a question of how a reaction network responds to a small change in kinetics along the lines of Eqn.~\ref{eqn:nuGeneral}. The small change represents the small structural difference between the right and wrong substrate. Formally, consider a reaction network with kinetic constants $\mathbb{W}_{ij}$ connecting intermediate states $i,j = 1, \ldots, N$ of an enzyme-substrate complex, described by a master equation as in Eqn.\ref{eqn:masterEqn}. Typically, we identify one of these states, say $i=0$ as the `reactants' state where the enzyme and substrate are unbound and another state, say $i = N$, as the `products' state where the substrate has been irreversibly processed to a final product, releasing the enzyme \cite{Hopfield:1974uo,Qian:2008wl}. (For an alternative proofreading model that leverages multiple inequivalent `reactants' states, see \cite{Hopfield:1980vm}.)

Hopfield and Ninio\cite{Hopfield:1974uo,Ninio:1975vv} realized that such discrimination $\nu$ is fundamentally limited at thermal equilibrium. In fact, at equilibrium, $\nu$ is constrained to be $1$ because the occupancy of state $p(E)$ is given by the  Boltzmann distribution $e^{-\beta E}$ and is completely independent of any details of the kinetics.

Hopfield and Ninio developed a non-equilibrium model based on intuition about discard pathways that repeatedly disrupt the reaction and allow discrimination to be enhanced. Such mechanisms are powered by ATP hydrolysis or other equivalent energy source that breaks detailed balance\cite{Schnakenberg:1976wb}; without such an energy source, discard trajectories are equally likely to be used as production pathways and would undo the error correcting properties. The idea of discard pathways was generalized shortly to allow arbitrarily high discrimination \cite{Ehrenberg:1980tv}, albeit at a very high cost of ATP turnover. 

Proofreading has been established in protein synthesis by ribosomes, DNA polymerase activity, charging of tRNA molecules and argued to play a role in immune receptor \cite{Hlavacek:2001iu,McKeithan:1995wq,Goldstein:2004ch,Francois:2013gx} ,  protein synthesis \cite{Blanchard:2004ce,Johansson:2008tz}, recA binding to DNA \cite{BarZiv:2002jd,Murugan:2012dz} and many other biochemical processes \cite{Yan:1999kx,AzizSancar:2004et}.
Proofreading and biological applications have been covered in numerous excellent reviews (e.g., \cite{Qian:2007gb,Goldstein:2004ch,Johansson:2008tz}). Here, we review only aspects relevant to trade-offs and design principles in generic and large proofreading networks;  thermodynamic effects and the existence of non-equilibrium phases are most relevant to the behavior of such large networks.

%In this review, we will use this interpretation of proofreading as an enhanced sensitivity to energy changes to connect to other biological mechanisms such as ultra-sensitivity and adaptation; such mechanisms provide enhanced or suppressed responses to other changes in kinetics due to substrate concentration changes or molecular binding events.
 
%These works have studied the resulting energy-error trade-offs for such mechanisms. 
 
\subsection{Design principles and types of proofreading networks}
The original proposal of Hopfield/Ninio was based on an intuition of discards. Generalizing \cite{Murugan:2014uq} to arbitrary networks without any special structure, proofreading requires the kinetics to favor a class of trajectories that bears out this intuition of discards. In a simplified picture, we assume that one preferred pathway in the network is chosen as the only path through which the reaction is completed; we call this path the reaction coordinate. (In reality, the reaction coordinate might consist of multiple allowed pathways, reduced down to a single effective coordinate using standard methods \cite{Li:2014tk})

\textit{Catastrophes}: Hopfield's and Ninio's (suitably generalized) intuition is that this reaction coordinate should have many discard pathways branching off it, with kinetics biased towards taking these pathways. When the system takes one of these discard pathways, we call that event a \textit{`catastrophe'} since progress along the reaction coordinate has been interrupted. (See Fig.~\ref{fig:JSP_discard_pathways}.)

\textit{Rescues}: Discard pathways necessarily undo progress along the reaction coordinate but they do not have to return the system back to the initial state; instead the system might be ``rescued'' to a point along the reaction coordinate upstream of the catastrophe, creating only a partial reset of the reaction.
 
 The possibility of rescues, i.e., avoiding complete resets after every catastrophe, is central to dynamical phase coexistence and the perspective of this review. Hence we show three networks with distinct rescue behaviors in Fig.~\ref{fig:JSP_network_examples} for purposes of pedagogy and to help understand the large body of network models in the literature. 
%This pathway should have many discard pathways branching off it, with kinetics biased towards taking these pathways. However, these discard pathways do not necessarily have to return the system back to the initial state; instead the discard events need to simply reset the reaction to an earlier state along the trajectory. 

%The large body of proofreading (and related driven networks such as for ultrasenstivity) in the literature can be collected into three classes:
%\textbf{More attractive names for these kinds of networks?}
\begin{enumerate}
\item[A)] No rescues: 
The networks of Fig.~\ref{fig:JSP_network_examples}A not allow for any possibility of rescue. All catastrophes take the system back to the origin. As argued below, these networks cannot support multiple dynamical phases and hence have a trade-off curve resembling Fig.~\ref{fig:JSP_Tradeoffs}A. Such models \cite{Ehrenberg:1980tv,Sontag:2001we,McKeithan:1995wq,Munsky:2009kp,DeRonde:2008vo,Goldstein:2004ch,Rabinowitz:1996wr,MacGlashan:2001gh,Hlavacek:2001iu} have attracted bulk of the attention, partly owing to early applications to immunology. It must be emphasized that while these models are easiest to analyze, there is little experimental evidence to favor these models over those with the possibility of rescue.
\item[B)] Variable range rescues:   We use the `ladder' network \cite{Ehrenberg:1980tv,Murugan:2014uq,Murugan:2012dz,Tu:2008vn,Mehta:2012ji} shown in Fig.~\ref{fig:JSP_network_examples}B as a stand-in for networks which rescue the system after varying amounts of backtracking following a catastrophe. For the ladder network shown, the length of backtracking prior to rescue is distributed exponentially from the point of catastrophe.
\item[C)] Short range rescues: An extreme case of networks that favor rescues is shown in Fig.~\ref{fig:JSP_network_examples}C where each catastrophe results in a short-ranged deterministic rescue. While these networks support multiple dynamical phases\cite{Vaikuntanathan2014} like those in (B), we use such networks as a pedagogical example to quantitatively explore phases in this review.
\end{enumerate}

Of course, complex networks can contain combinations of all three of the above archetypes. With disordered networks with varying rates of catastrophes and rescues, novel phenomena\cite{SINAI:1993ui} related to random walks in disordered potentials  have been seen \cite{Murugan:2012dz} but has not been fully investigated. For pedagogical reasons, here we focus on networks like those shown in Fig.~\ref{fig:JSP_network_examples} whose rates of catastrophes and rescues do not vary across the network.

\subsection{Trade-offs}
Networks of any kind shown in Fig.~\ref{fig:JSP_network_examples} can proofread. However the trade-off between error and dissipation or time cost is determined by the balance of catastrophes and rescues.  

The question of trade-offs in proofreading was first studied by Bennett \cite{Bennett:1979tb}; working with small networks, he derived equations for the minimum dissipation needed to achieve a given error rate.
For recent work on trade-offs in proofreading (including alternate trade-offs between chemical potential and error rate)  and related works on finite networks, see \cite{Munsky:2009kp,Johansson:2008tz,Ehrenberg:1980tv,Qian:2006dl}. Here, we focus on large networks which opens up qualitatively new phenomena related to phase coexistence.
%Two aspects of these discard pathways are relevant for the purposes of error correction:
%1. catastrophe probability distribution - the probability with which the system will take a discard pathway at a given position ‘x’ along the reaction coordinate. 

%2. rescue probability distribution - the probability of arriving back at a position ‘y’ along the reaction coordinate, given a catastrophe at position ‘x’. 

%General networks will have catastrophe and rescue properties that vary along the reaction coordinate. 

%Thus, a small difference in catastrophe rates for R and W is greatly enhanced into a large difference in reaching the final state due to the collective behavior of many $n$ discard pathways. 

Since the right and wrong substrate bind to the enzyme with slightly different energies, the catastrophe and rescue probabilities for the right and wrong substrate differ by a small amount. In the simplest model (see Fig.~\ref{fig:JSP_discard_pathways}), we assume that difference is entirely in the catastrophe rates and the kinetic constant $d$ on the discard pathway differs for the two substrates. As a result, the probabilities of having a catastrophe (i.e., taking the discard pathway) at any particular site for the right and wrong substrates are given by,
\begin{equation}
p^R_{cat} =  \frac{d}{d+f}, p^W_{cat} = \frac{d e^{\beta E}}{d e^{\beta E}+f}
\label{eqn:pcat}
\end{equation}
where the binding energy of $EW$ is lower than $ER$ by an energy $E$.

1. \textit{Both substrates localized at reactants.} 
%lowest error requires very frequent catastrophes - both are localized at origin. Fig 4a.
%both cats are very high.. requires f to be very small. 

The limit of lowest possible error (and associated very high costs) is easy to understand and accounts for much of the proofreading literature. In this limit, we ignore the possibility of rescues (either because the network cannot support them or because the rate of rescues is low)
and assume that catastrophes are very frequent for both substrates (see inset in Fig.~\ref{fig:JSP_discard_pathways}),
$$f \ll d < d e^{\beta E}.$$
The error rate $\eta \sim (p_W/p_R)^n$ goes to its lowest possible value, 
\begin{equation}
\eta \to e^{-n \beta E}
\end{equation}
where $n$ is the number of discard pathways along the reaction coordinate where catastrophes can take place. However, the time to complete the reaction with the right substrate is exponentially large in $n$,
$$T \sim (1-p^R_{cat})^{-n} \sim \left(\frac{d+f}{f}\right)^n.$$ 

\begin{figure}
 \centering
 \includegraphics[width=0.7\linewidth]{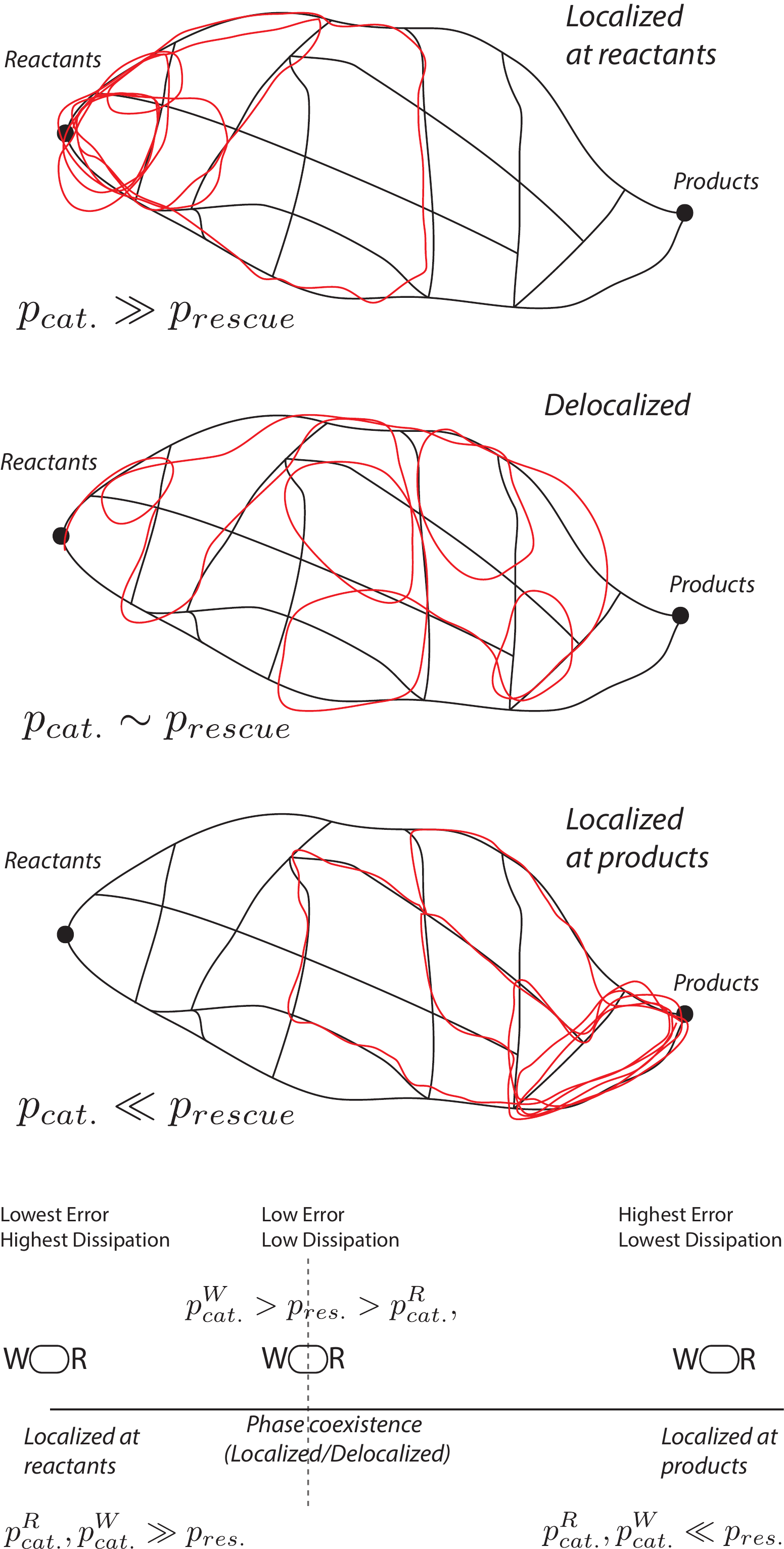}
 \caption{Typical trajectories in a dynamical phase are determined by the balance of catastrophe and rescue rates. (A,C) If either the catastrophe or rescue rate is much higher than the other, the system is localized at one end of the network. The state occupancy decays exponentially away from the localization site. Typical trajectories involve small cycles passing through such a site. (B) If catastrophe and rescue rates are comparable, the system is de-localized and explores the whole network. (D) If substrates $R$ and $W$ are both localized at the reactants or products end of the network, we find minimal error but at very high cost or vice-versa. Instead, near the co-existence point, we find much reduced dissipation and only a small increase in error rate (natural compromise point of Fig.~\ref{fig:JSP_Tradeoffs}B). 
In this regime, the small difference in catastrophe rates for $R$ and $W$ leads to dramatically different behavior for the two substrates; $R$ is (nearly) localized at products while $W$ is (nearly) localized at reactants.   \label{fig:JSP_LocalDelocal}}
 \end{figure}

%If both catastrophe rates are much higher than the rescue rates, both substrates are localized near reactants (Fig  A) and we can immediately estimate 
%the error rate would be
In this limit, both substrates spend their time `localized' near the reactants end of the network since catastrophes frequently reset both of their reactions. See Fig.~\ref{fig:JSP_LocalDelocal}A. Hence occupancy $\psi$ of states in Fig.~\ref{fig:JSP_discard_pathways} decays exponentially from the reactants end of the network even for the right substrate. This regime is supported by all the networks shown in Fig.~\ref{fig:JSP_network_examples}.

%We say that both substrates as ‘localized’ near the reactants end of the network .

%The above picture explains how to achieve the smallest possible error rate; however, this limit requires a high catastrophe rate for both right and wrong substrates and hence the enzyme-substrate complex spends most of its time near the reactants part of the network even when the substrate is the right one. As a result, the cost of such a low error rate ($e^{-n \Delta}$) is very high (possibly infinite) dissipation and time. 

2. 
\textit{Both substrates localized at products.} 
Alternatively, if discards are unlikely for both substrates, i.e. if $f \gg d e^{\beta E} > d$ ,  the catastrophe rates $p_{cat}^R,p_{cat}^W$ are very small compared to rescue rates. Both substrates, on average, make linear progress towards the reactants end of the network. In fact, the reactions with both substrates are now effectively localized at the products end of the network (see Fig.~\ref{fig:JSP_LocalDelocal}C) and when reactants are released, the reaction is completed quickly. However, while the reaction time is greatly reduced to linear in $n$, the error rate is back to the (high) equilibrium level, $$T \sim n, \eta \sim e^{-\beta E}.$$
%high catastrophe rate for both right and wrong substrates and hence the enzyme-substrate complex spends most of its time near the reactants part of the network even when the substrate is the right one. No dissipation, high error.

3. \textit{Phase coexistence}
If the reaction network can support rescues sufficiently (as in Fig.~\ref{fig:JSP_network_examples} B or C but not A), the transition between these two regimes involves a surprisingly favorable regime. When the catastrophe rate becomes comparable to the rescue rate, the system executes trajectories that are delocalized (see Fig.~\ref{fig:JSP_LocalDelocal}B) and explore the whole network. In fact, as we show in the next section, the trajectories are a mix of localized and delocalized trajectories and when the catastrophe rate matches the rescue rate, we have coexistence between localized and delocalized trajectories. 

However, the right substrate $R$ has a lower catastrophe rate than $W$ (see Eqn.~\ref{eqn:pcat}); hence $R$ is closer to being localized at the products (regime (2)) while $W$ is closer to being localized at the reactants (regime (1)). As a result, we find very fast reaction completion times for $R$ while still greatly suppressing the reaction rate for $W$,
$$ T \sim n, \quad  \eta \sim e^{-n \beta E'}$$
where $0 < E' < E$. Hence at coexistence, we find a enormous speed-up of the reaction at only a small cost in error rate, (shown as the natural compromise point in Fig.~\ref{fig:JSP_Tradeoffs} B).

Intuitively, we find that the two substrates can be localized at different ends despite having a small difference in kinetics. 
Note that at phase coexistence, we do pay a price in the form of large variance in reaction completion times for the right substrate. However, the mean completion time near phase coexistence is significantly lower than in the regime with lowest error (i.e., when both substrates are localized at reactants).

In this perspective on proofreading, we are using the delocalized phase as an inevitable intermediate in order to switch quickly from one localized phase to another. In other systems, being able to switch easily from localized to the delocalized phase itself might have function consequences.

%Old stuff:

%It is easy to see that as $\eta \to \eta_{min} = e^{-n \Delta}$ , we need $ f /d \to 0$ and hence $T \sim (f/d)^n \to \inf$, exponentially in the error rate. In this limit, we can write, $T \sim (1/(\eta - \eta_{min})^n$.

\section{Localized and delocalized phases.}

The calculations presented in the previous section demonstrate that simple models of kinetic proofreading and microtubule growth dynamics exhibit multiple dynamical regimes~\cite{Murugan:2012dz}. The transition between regimes is sharp and has a thermodynamic quality. The language of large deviation theory~\cite{Touchette2009} can be used to formally understand the transition between different kinetic regimes and  to anticipate ways in which the transition can be generalized. Indeed, dynamic heterogeneity and intermittency have been observed in many other biophysical systems including the dynamics of kinetic proofreading by RNA polymerase on m RNA strands~\cite{Depken2013}, and the dynamics of gene regulatory networks~\cite{Suel2006}.

We begin by reviewing some simple models of non equilibrium dynamics~\cite{Vaikuntanathan2014}. The strategy adopted in these studies revolves around computing the cumulant generating functions and the large deviation functions for entropy production in these networks~\cite{Lebowitz1999}. From these calculations it is apparent that the trajectories produced in these simple models can be separated into two classes: one class of trajectories is delocalized throughout the network, and a second class is localized around a few points in the network. These classes of trajectories are separated by a dynamical phase transition. 
The calculations show that a class of non equilibrium systems can, under very general conditions, support two different dynamical modes. Similar inferences can be drawn by considering the statistics of other path dependent observables such as first passage times.

\subsection{ Large deviation theory and importance sampling in trajectory space}

The probabily of observing a particular value of entropy production, $\omega$, in a trajectory of length $\tau\gg1 $ obeys a large deviation principle~\cite{Lebowitz1999},
\begin{equation}
P(\omega) \approx e^{-\tau I(\sigma)}\,
\label{eq:largedeviationprinciple}
\end{equation}
where $\sigma\equiv \omega/\tau$ is the entropy production rate, and $I(\sigma)$ is the large deviation rate function.
The large deviation rate function $I(\sigma)$ can be determined from direct simulation of $P(\omega)$, though large deviations are difficult to sample as $\tau$ grows.  
Alternatively the convex envelope of $I(\sigma)$ can be computed as the Legendre transform of the scaled cumulant generating function $\psi_\omega(\lambda)$~\cite{Touchette2009,Lebowitz1999}
\cite{Touchette2009, Lebowitz1999},
\begin{equation}
\psi_\omega(\lambda) = \lim_{\tau \rightarrow \infty} \frac{1}{\tau} \ln \left<e^{-\lambda \omega}\right>,
\label{eq:cgf}
\end{equation}
where the expectation value is taken over trajectories initialized in the steady state distribution.

Following the general framework laid out by Lebowitz and Spohn~\cite{Lebowitz1999}, the cumulant generating function $\psi_\omega(\lambda)$ can be computed as the maximum eigenvalue of a matrix operator, $\mathbb{W}_\omega(\lambda)$, which is simply related to $\mathbb{W}$, the transition matrix for the kinetic network~\cite{Lebowitz1999}.  
Specifically the $ij$ matrix element is given by
\begin{equation}
\mathbb{W}_\omega(\lambda)_{ij} = \left(1 - \delta_{ij}\right)\mathbb{W}_{ij}^\lambda \mathbb{W}_{ji}^{1-\lambda} + \delta_{ij} \mathbb{W}_{ij}.
\label{eq:Woperatordef}
\end{equation}
By solving for the eigenspectrum of $\mathbb{W}_\omega(\lambda)$ the cumulant generating function $\psi_\omega(\lambda)$ and therefore the large deviation function $I(\sigma)$ can be computed via a Legendre transformation. We also note that the fluctuation theorem implies a symmetry in the scaled cumulant generating function, $\psi_\omega(\lambda)=\psi_\omega(1-\lambda)$~\cite{Lebowitz1999}. 

The eigenvectors of $\mathbb{W}_\omega(\lambda)$ also hold special significance~\cite{Chetrite2013}\cite{Jack2010}. They reflect the character of the dominant trajectories in an ensemble in which trajectories have been importance sampled~\cite{Dellago1998} according to the bias $\exp(-\lambda \omega)$ where $\omega$ is the entropy produced along the trajectory. Indeed, when $\lambda=0$, the biasing function is unity and left eigenvector of the $\mathbb{W}_\omega(\lambda)$  is simply proportional to the steady state probabilities.

Discontinuities in the first derivative in the cumulant generating function signal the presence of dynamically heterogeneity and intermittency in the dynamics. 
Specifically, a discontinuity in the cumulant generating function implies the presence of two dynamic phases. The corresponding large deviation rate function has a tie line connecting average entropy production rate of the two phases ~\cite{Touchette2009,Garrahan2007}. Dynamical phases in glassy materials,\cite{Garrahan2007}, protein folding dynamics~\cite{Weber2013}, and models of driven systems~\cite{Hurtado2011v2} have been discovered using such approaches.

%Similar approaches have been used to reveal dynamic heterogeneity and intermittency in glasses[Garrahan], models of driven systems [Garrido], and more recently in protein folding dynamics[Jack, Pande]. For instance, by computing the statistics of trajectory dependent quantities, Jack et al found 

\subsection{Dynamical heterogeneity in biochemical networks} 
A natural question is whether the formalism of large deviation theory and statistical mechanics in the space of trajectories can be used to understand the presence of multiple classes of trajectories in kinetic proofreading and microtubule growth models. Indeed, analysis of the cumulant generating functions for first passage times (and other observables that reflect the transport properties of trajectories) have yielded fruitful results in models of kinetic proofreading~\cite{Murugan:2012dz}.  

\begin{figure}[h!]
\centering
\includegraphics[width=0.90\linewidth,angle=0]{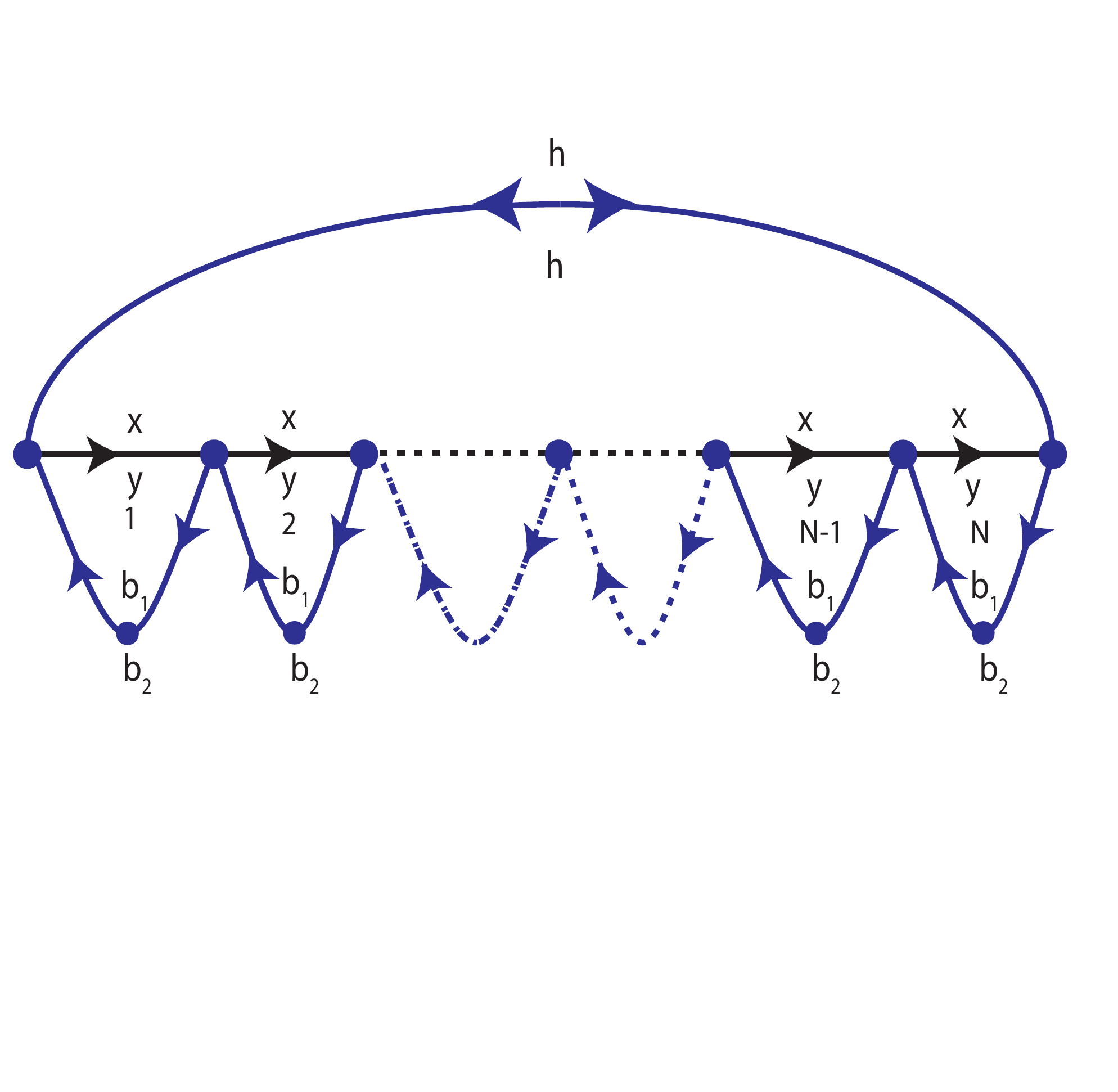}
\caption{A network similar to one the depicted in Fig.~\ref{fig:JSP_network_examples} (C) with short ranged rescues and catastrophes. The $h$ link connects up the ends of the network and is mean to model the rates associated with the cycling between the reactant and product states.}
\label{fig:trianglenetwork}
\end{figure}

Consider the network shown in Fig.~\ref{fig:trianglenetwork}. It consists of triangular motifs which are replicated around in a chain. 
The symmetry is broken with  a $h$-link (see Fig.~\ref{fig:trianglenetwork}). 
Each triangular motif asymmetric links that ensures cycling currents on average. This system of connected triangular motifs is similar to the kinetic proofreading network in Fig.~\ref{fig:JSP_network_examples} (C). Each triangle segment can be thought of as offering two different pathways, one of which is catastrophe followed by an immediate rescue, to go between two states. The $h$ link in this network mimics the rate at which the polymerase enzyme cycles through the proofreading process. When used as a model of kinetic proofreading, the asymmetry in the rates can either be due to energetic or kinetic considerations~\cite{Sartori:2013fv}.

\begin{figure}[h!]
\centering
\includegraphics[width=0.90\linewidth,angle=0]{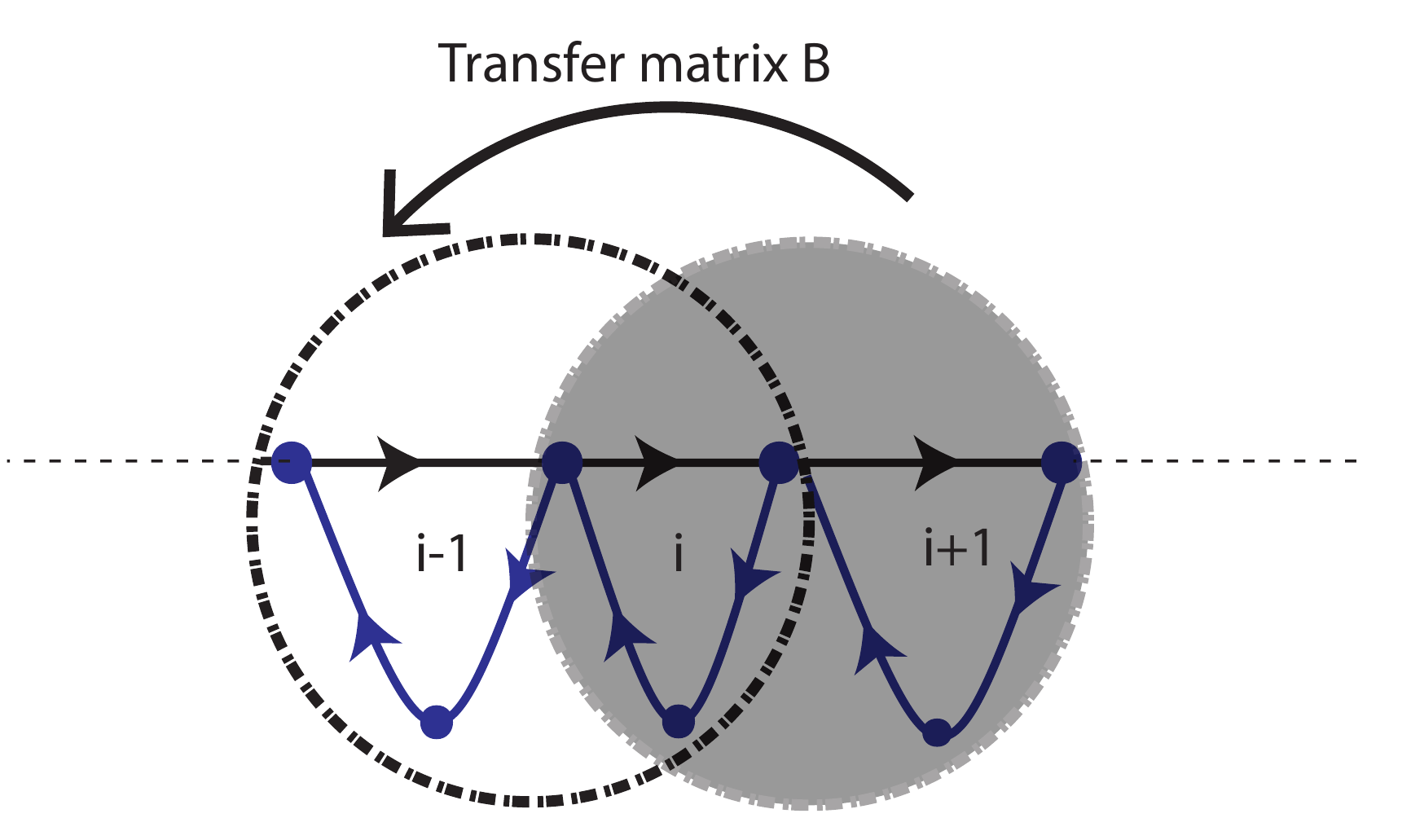}
\caption{The largest eigenvalue and eigenvector of the titled matrix, Eq.~\ref{eq:Woperatordef} are computed by expressing the eigenvalue problem $\mathbb{W}_\omega(\lambda) \hat f_i =\psi_\omega(\lambda)\hat f_i $ using transfer matrices. For elements that are not adjacent the heterogenous link, a transfer matrix $B$ connects elements of the eigenvector inside the shaded circle to elements of the eigenvector inside the open circle. Similar transfer matrices, $A_1$ and $A_2$ are defined for elements adjacent and spanning the heterogeneous link.}
\label{fig:triangletransfer}
\end{figure}

\begin{figure}[h!]
\centering
\includegraphics[width=0.90\linewidth,angle=0]{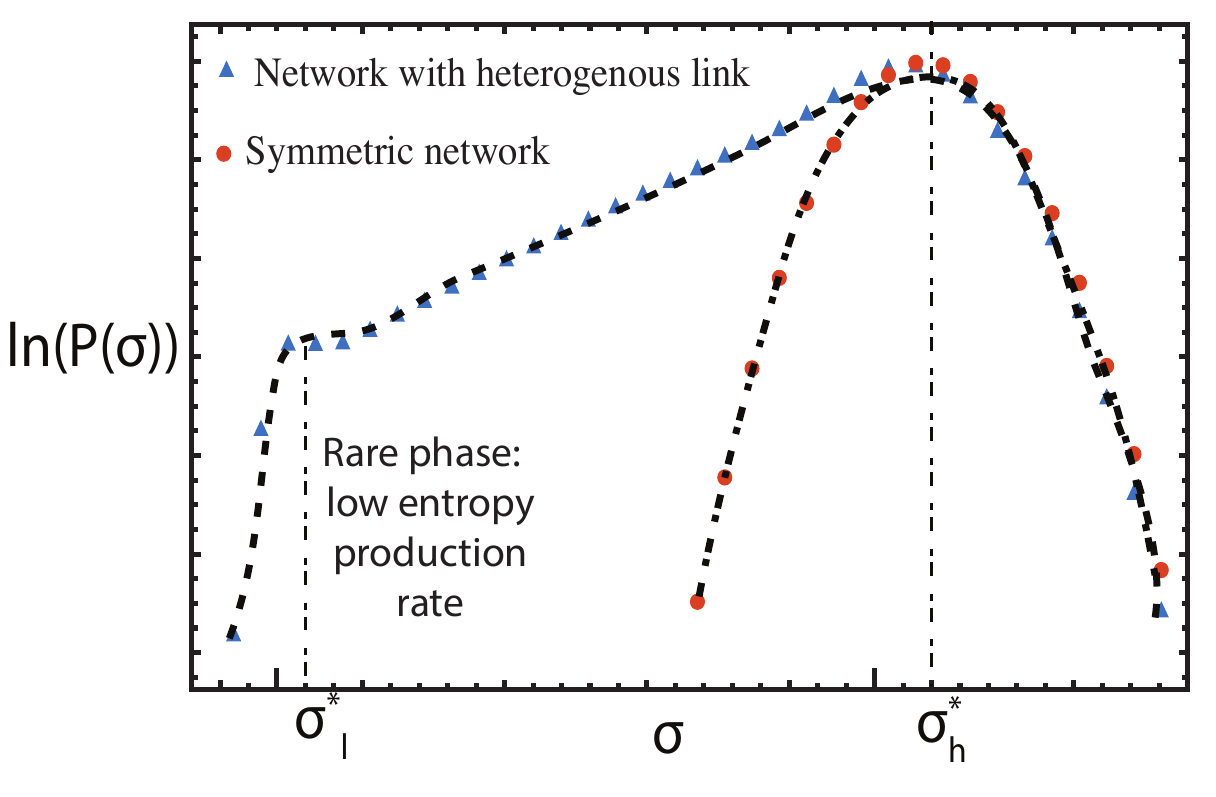}
\caption{The probability distribution associated with observing an entropy production rate $\sigma$ in the network. The probability distribution of the network with the heterogenous link has a second rare phase which produces entropy at a lower rate $\sigma_l^*$. The probability distribution is localized around the heterogeneity when the system is in this dynamical this phase. By comparison, the probability distribution of the symmetric network has a single dynamical phase producing entropy at a rate $\sigma_h^*$}
\label{fig:LDF}
\end{figure}

\begin{figure}[h!]
\centering
\includegraphics[width=\linewidth,angle=0]{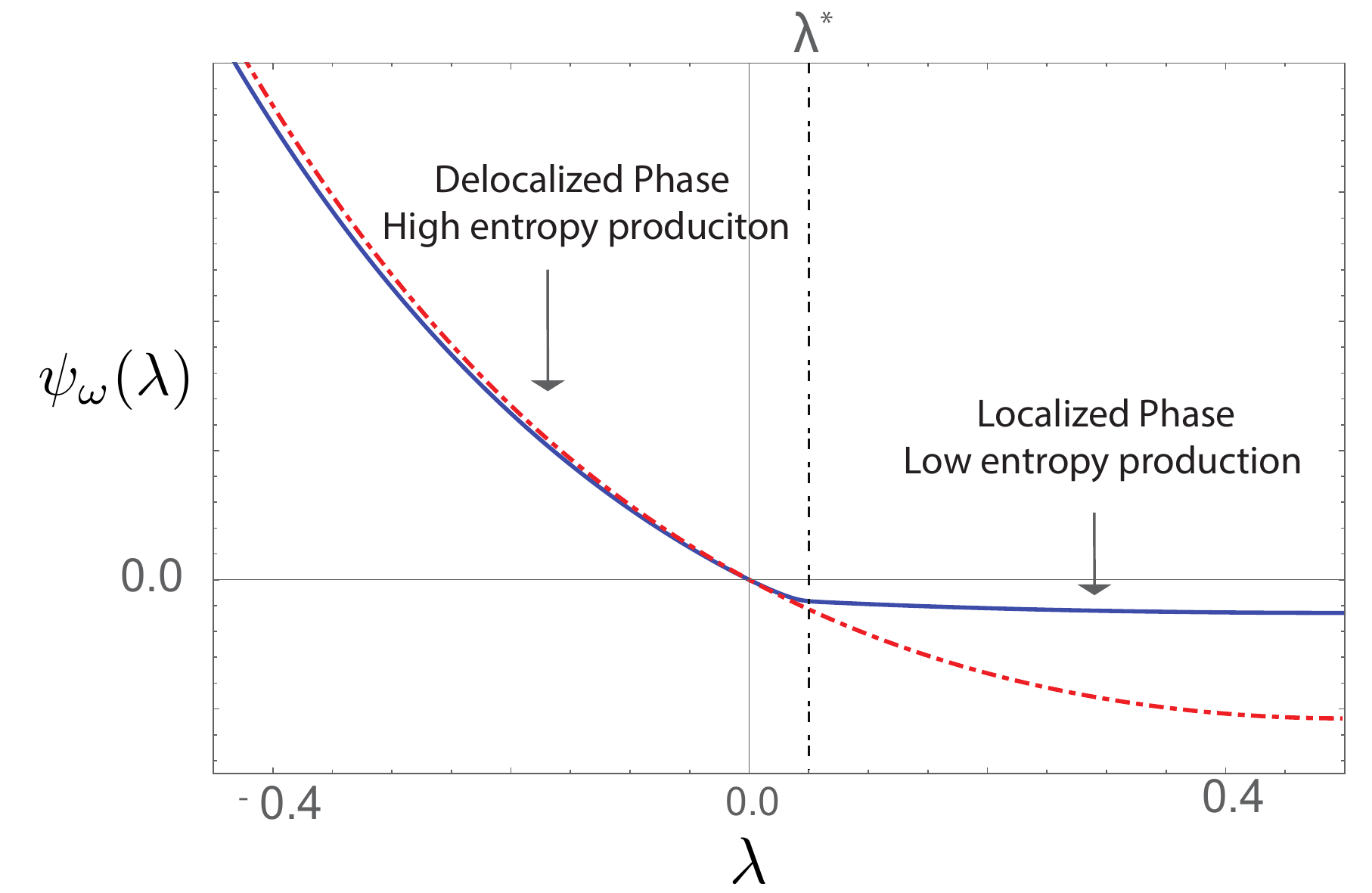}
\caption{The cumulant generating function for the triangle network in Fig.~\ref{fig:trianglenetwork} for a particular set of rates (Solid line) has a kink or a discontinuity in its first derivative. In comparison, the cumulant generating function of translationally symmetric variant (dotted line) has no such singularity. The slope of the cumulant generating function is related to the rate of entropy production.}
\label{fig:kink}
\end{figure}

The cumulant generating functions and the large deviation functions for this simple network reveal interesting features. Analysis of the large deviation function reveals contributions from two classes of trajectories. One class of trajectories is delocalized throughout the entire network. The second rare class of trajectories is localized near heterogenous link (Fig.~\ref{fig:LDF}). The delocalized trajectories consume energy at a higher rate. Indeed, numerical simulations reveal that the cumulant generating function exhibits a cups or a singularity in the slope at particular values of $\lambda$ (Fig.~\ref{fig:kink}). These cusps signal a dynamical phase transition. 

To clarify the nature of these phases we review the calculation of the cumulant generating function. 
First, consider a network in which $N$ triangular motif are connected together in a ring configuration. This network has translational symmetry. 
The tilted matrix of the translationally symmetric variant of the triangle network (Fig.~\ref{fig:trianglenetwork}), $\mathbb{W}^{\rm ts}_\omega(\lambda)$, can be diagonalized by a discrete Fourier transform. The largest eigenvalue of this matrix, or the scaled cumulant generating function  $\psi^{\rm ts}_\omega(\lambda)$, is smooth.  The rate function resulting from the Legendre transform of $\psi^{\rm ts}_\omega(\lambda)$ is a simple convex function peaked around the average entropy production rate. With no broken symmetry there is only one dynamical state.

Networks with the $h$ link support a second phase. The cumulant generating function for the function with the $h$ link can be computed using a novel perturbation theory. We review this approach below. Let $\hat f_i$ denote the elements of the right eigenvector corresponding to the largest eigenvalue of $\mathbb{W}_\omega(\lambda)$. The eigenvalue problem $\mathbb{W}_\omega(\lambda) \hat f_i =\psi_\omega(\lambda)$ can be recast in terms of transfer matrices~\cite{Derrida1987,Vaikuntanathan2014}. Specifically, a transfer matrix $B[\psi_\omega(\lambda)]$ can be constructed such that it maps certain elements of the eigenvector inside the shaded circle in Fig.~\ref{fig:triangletransfer} to the elements inside open circle. This transfer matrix acts on elements away from the heterogenous link. Similar transfer matrices $A_1$ and $A_2$ can be constructed across the heterogenous link.  Using the notation $\hat f_i\,,\hat f_{i+1}$ to denote the collection of elements inside the shaded and the open circle respectively in Fig.~\ref{fig:triangletransfer}, and performing the transfer matrix operation around the ring,  the eigenvalue problem can be recast into 
\begin{equation}
\hat f_1 = B^{N-2} A_2 A_1 \hat f_1
\label{eq:BAAequality}
\end{equation}
The eigenvalues of  $\mathbb{W}_\omega(\lambda)$, in particular the largest eigenvalue $\psi_\omega(\lambda)$, have to satisfy Eq.~\eqref{eq:BAAequality}, which requires that that the matrix $B^{N-2} A_2 A_1$ have an eigenvalue $1$. 
The corresponding eigenvector can be used to obtain the elements of the largest eigenvector.

The eigenvalue problem in Eq.~\ref{eq:BAAequality} cannot be immediately solved since the transfer matrices themselves depend on the cumulant generating function $\psi_\omega(\lambda)$~\cite{Vaikuntanathan2014}. 
Fortunately, in the large $N$ limit, the system with the $h$-link can be analytically solved using a perturbation theory around the solution of the fully periodic network. Specifically, the following perturbation ansatz can be proposed. 
\begin{equation}
\psi_\omega(\lambda) = \psi^\text{p}_\omega(\lambda) + \frac{\gamma}{N} + \mathcal{O}\left(\frac{1}{N^2}\right)\, 
\label{eq:pertubativeexpansion}
\end{equation}

The equation Eq.~\eqref{eq:pertubativeexpansion} provides a perturbative solution to the aperiodic system. The perturbation theory is self consistent and breaks down when the vale of $\gamma$ diverges. This breakdown happens as $\lambda$ approaches a particular value $\lambda^*$. For $\lambda \geq \lambda^*$, the perturbative approach is not valid. 
Thus, in the large $N$ limit, the broken symmetry network behaves exactly like the periodic network up to that value $\lambda^*$, at which point it deviates markedly (Fig.~\ref{fig:kink}).

When $\lambda < \lambda^*$, the components of the eigenvector, $f_n$, resemble those of the translationally symmetric network and have a predominantly delocalized character, $\ln f_n\sim n/N$. Entropy is dissipated as the system goes through global cycles in the delocalized state. 

When $\gamma$ diverges as $\lambda \rightarrow \lambda^*$, the translationally symmetric network no longer offers a good perturbation basis. Rather, the eigenvectors are now localized around the inhomogeneity of the network $\ln f_n\sim -\epsilon n$. The competition between the rescue/catastrophes and the heterogeneities in the network couple to stabilize a localized mode. Entropy is generated by current fluxes in local cycles around the heterogeneity. The entropy generation rate in the localized and delocalized phases are markedly different. In the setup reviewed here, the delocalized phase consumes energy at a higher rate as the system cycles the entire network. 

The relative fractions of localized and delocalized phases is related to the location of the kink, $\lambda^*$. 
Specifically, the slope of the tie line in the large deviation rate function connecting the entropy generation rates characteristic of the localized and delocalized phases is equal to $\lambda^*$,
\begin{equation}
\lambda^*=(I(\sigma^*_h)-I(\sigma^*_l))/(\sigma^*_h-\sigma^*_l)\,,
\end{equation}
where $\sigma^*_h$ is the entropy generation rate characteristic of the delocalized phase and $\sigma^*_l$ is the entropy generation rate characteristic of the localized phase. The localized and delocalized phases \textit{coexist} when kinetic constants are chosen so that $\lambda^*=0$.

The transfer matrix framework reviewed here is applicable more generally. Indeed, we note that similar transfer matrix techniques have been used to study single particle localization in strips of disordered media~\cite{Derrida1987}. This analogy suggests that localized and delocalized phases should be present in a larger class of biochemical networks including for example the network with finite range rescues and catastrophes as described in Fig.~\ref{fig:JSP_network_examples} (B). 

While the networks reviewed here support the delocalized phase in its steady state \textemdash the localized state occurs in a rare sub ensemble of trajectories \textemdash kinetic rates can also be constructed so that system is localized in the steady state.
The localized subclass of trajectories can be made more likely by sampling trajectories according to the biasing function $\exp(-\lambda \omega)$. Applying a bias equal to $\exp(-\lambda^* \omega)$ makes the likelihood of the two phases equal. Further, using the Doob h-transform~\cite{Chetrite2013}, it is in fact possible to construct Markov processes that generate ensembles equivalent to the ensemble of trajectories generated through the application of biasing techniques. This framework allows for the design of networks which can be tuned to support  both localized and delocalized regimes.

The existence of multiple dynamical phases implies that such kinetic networks can be tuned to serve different purposes. 
%The existence of multiple dynamical phases in these simple kinetic networks has many interesting implications. 
Networks in their localized state illustrate how non equilibrium dynamics can be used to achieve control \textemdash \ the system mainly samples states near the $h$-link \textemdash \ 
at the cost of energy dissipation. 
Indeed, as noted previously networks similar to that in Fig.~\ref{fig:trianglenetwork} are used in modes of kinetic proofreading~\cite{Murugan:2014uq} to achieve discrimination between two macromolecular states. Moreover, the analytical framework described above can be used to estimate the energy requirements (measured for example by the rate of entropy dissipation) in these and other similar feedback and control processes~\cite{Lan:2012in}.  
The steady state of the network can either be localized near the states $1$ or near the state $N$ depending on kinetic constants in the network. The two localized edge states map onto the bound and unbound modes in the context of microtubule growth, and to localized modes at the reactants or products in the context of kinetic proofreading (Fig.~\ref{fig:JSP_network_examples}(C))

The transition between these two states is sharp when $N\gg1$. A network poised localized near the reactant nodes but poised close to a dynamical phase transition can be particularly effective for kinetic proofreading. Such a network will have a high throughput due to its ability to utilize the high transport properties of the de-localized state. The trade-off between error rates and throughput rates can hence be optimized by tuning a system close to a dynamic transition. As discussed below, for microtubule growth, the dynamics close to phase transition point provides an optimal search strategy. 
%Finally, the connections reviewed in this section also suggest a general mechanism for intermittency in biophysical systems. 

 %Our analysis show that such networks can generically support a delocalized phase allows the system to more effectively sample its state space.  
%We anticipate that similar calculations will become useful in identifying dynamic regimes in nonequilibrium self assembly~\cite{Whitelam2012}, and non equilibrium polymer growth problems

%\begin{figure}[tbp]
%\centering
%\includegraphics[width=0.99\linewidth]{LadderNetwork_Edit4}
%\caption{Transitions between localized states can be mediated by non equilibrium forces.}
%\label{fig:network2}
%\end{figure}

\section{Implications for other biophysical processes: Open questions and future directions}

\subsection{Search problems in weak gradients}

%\textbf{Could make a distribution of reset lengths}
%Reset length = sudden ballistic traversal of reaction coordinate in a 'short time'.. 
%Define this. In the search problem, you walk and then fall back by some distance. Let's say you are restricted to small unit steps and then occasionally to large steps backwards..  but the net bias needs to be held fixed? 

\begin{figure}
\centering
\includegraphics[width=0.7\linewidth]{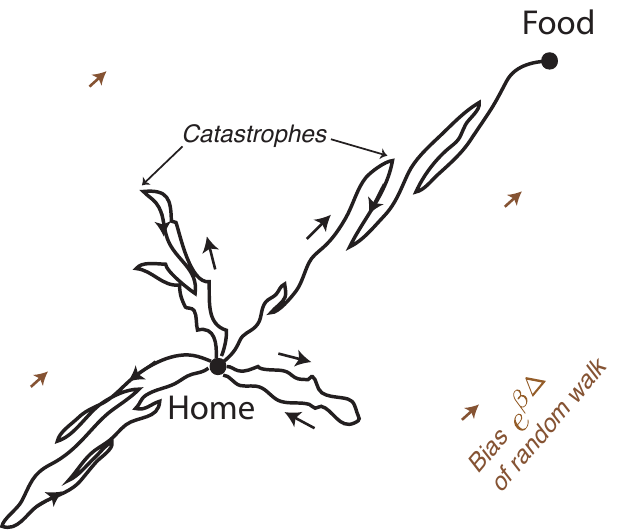}
\caption{Search for `Food' through a weakly biased random walk can be greatly enhanced by adopting trajectories with ``dynamic instability''. If such trajectories are posed close to the localization-delocalization transition, small differences in catastrophe rate in different parts of space will have magnified consequences on large scale motion; motion can be rapid in the direction of `Food' while still highly suppressed in other directions. Microtubules are thought to face such a search problem when locating centromeres during spindle formation; dynamic instability of microtubules, if tuned to coexistence of bounded and unbounded growth can greatly speed up such a search. \label{fig:JSP_Search}}
\end{figure}

The trajectory picture emphasized in this review also suggests new search strategies in uncertain environments. Adding an active non-equilibrium component to random walker's exploration statistics and tuning this non-equilibrium component to coexistence between localized and delocalized trajectories can greatly reduce the time to find a target.

Consider a random walker that leaves the `Home’ state (see Fig.~\ref{fig:JSP_Search}) and executes a random walk with fixed step size that is weakly biased towards a target (`Food'). (We denote this fixed small bias $e^{\beta \Delta}$ to connect to proofreading.) Such equilibrium exploration of the environment will result in a long time to find the food source, especially when `Food' is far relative to the step size of the random walker and the bias is small. 

Instead, consider ``active exploration'': after the random walker has randomly decided to take a step backwards, the walker can randomly decide to make the size of the backward motion large compared to the usual step size. Such a large step is allowed to reverse a finite fraction of the displacement from `Home'. The initiation and termination of such a reversal are analogous to catastrophes and rescues discussed earlier in the context of proofreading; in particular, the length of such reversals is related to the balance of catastrophe and rescue rates.

 ``Active exploration'' can be thought of as exploration with alternate statistics of returns `Home' but with no greater intrinsic bias towards `Food'. Nevertheless, search is greatly enhanced by tuning the active component of the exploration statistics to the localization-delocalization transition.
 
In such a regime, even if the bias is weak and the variations in catastrophe rates in space is small, the net effect of such a change on large scale motion can be dramatic. Trajectories that head in the wrong direction have catastrophic reversals at a rate that is just high enough to localize the walker at `Home'. On the other hand, trajectories in the right direction might begin similarly biased towards `Home' but have a falling catastrophe rate as they proceeds towards `Food'. Once the catastrophe rate falls below a threshold, the walker moves rapidly forward and finds the `Food'.  `Food' is effectively brought closer to the walker, positioned at a point where the catastrophe rate falls below a threshold.  %As a result, much like in proofreading, the time required to find the food source can be dramatically reduced. 

%Note that the size of the large backward step is not allowed to depend on the relative placement of the food (or equivalently the bias). Hence 

%If the size of reversals towards home is too large, the walker is localized at `Home'; `Food' is found only on an exponentially unlikely sojourn. 
%In some geometries related to microtubule spindle assembly, such a limit was argued to speed up search by avoiding long forays in the many wrong directions \cite{Holy:1994wf}. 

%If catastrophes are very frequent, the walker is localized at `Home’. Close to the transition, 

Microtubules are thought to face a similar problem in the search for chromosomes (or centromeres) during mitotic spindle assembly \cite{Holy:1994wf,Kirschner:2006us}. The framework of dynamical phase coexistence emphasized in this review suggests that the search for chromosomes is greatly sped up when microtubule growth is posed close to phase coexistence between `bounded' and `unbounded' growth, i.e., when the rate of catastrophes is comparable to the rate of rescues. In this limit, even a weak gradient of molecules that can modulate the catastrophe rate can have a dramatic impact on large scale motions. Evidence for such molecular mechanisms that can modulate catastrophe rates has emerged in recent years \cite{BowneAnderson:2013ik}.

\subsection{Other ways of encoding substrate differences}
Within the context of proofreading, recent work \cite{Sartori:2013fv,Rao:2015vy} has classified proofreading schemes by whether the difference between right and wrong substrates is encoded in differing binding energies (as in the original papers \cite{Hopfield:1974uo,Ninio:1975vv}) or in the heights of activation barriers (as in subsequent work by Bennett \cite{Bennett:1979tb}). The latter models can be described by Eqn.~\ref{eqn:nuE} if binding energy $E$ is replaced by the height of an activation barrier in the network. While small networks have been studied by \cite{Sartori:2013fv}, the response of large generic networks to changes in activation barriers remains unexplored. Many biochemical systems might well rely on such discrimination coded in activation barriers \cite{Ling:2007ep}.

\maketitle
\begin{figure}[h!]
\centering
\includegraphics[width=0.90\linewidth,angle=0]{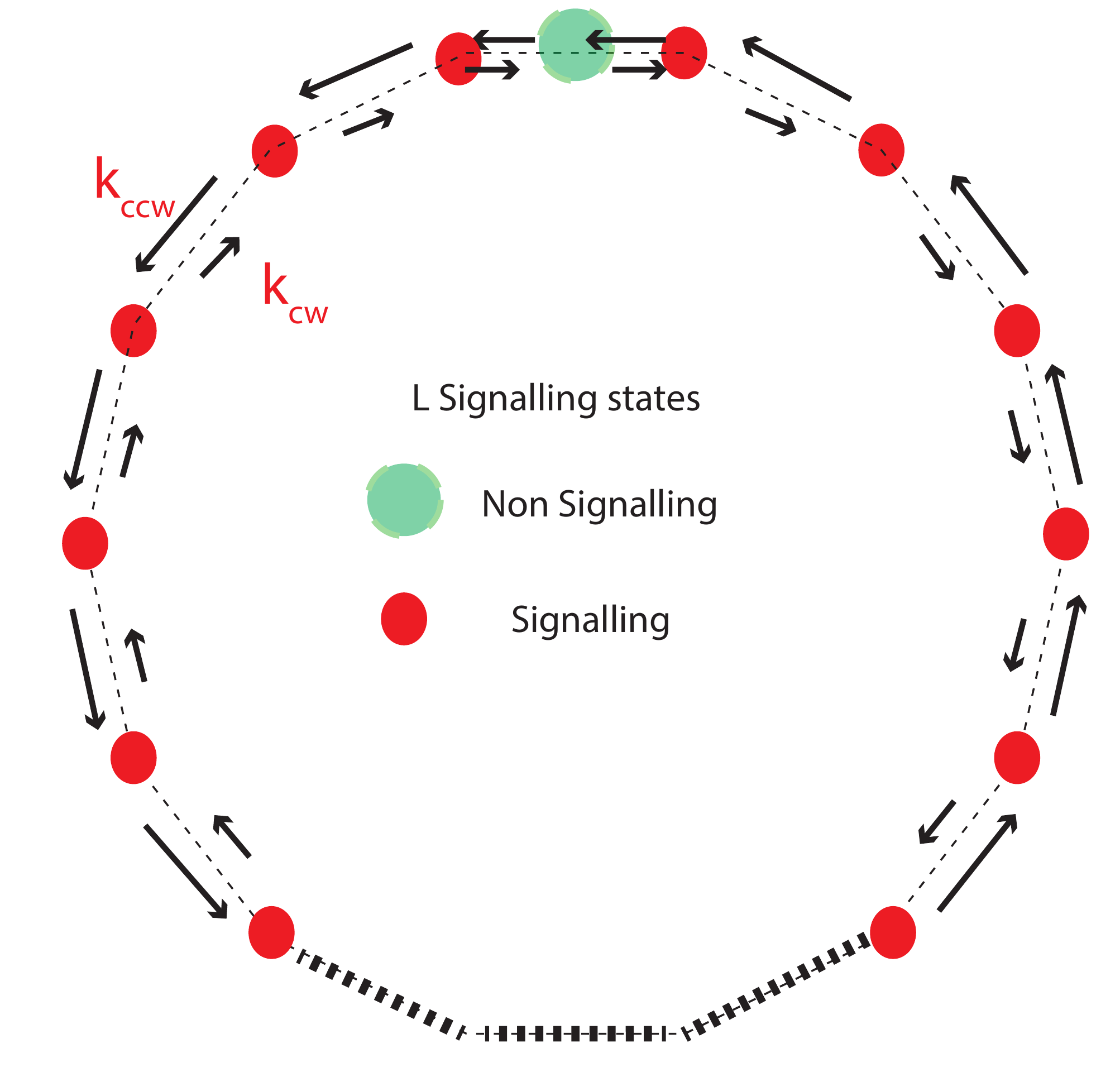}
\caption{Schematic of the network \cite{Lang:2014ir} used to model a signaling receptor that measures the concentration of an external ligand. Ligand binding promotes occupancy of the signaling states and ligand concentration is read out as the time $\tau_s$ spent in such signaling states. The receptor can measure concentrations accurately by expending free energy (e.g., coupling ATP hydrolysis along the ring shown) and reducing the variance of $\tau_s$. A particularly favorable regime of highly enhanced accuracy at a small free energy cost (in the sense of Fig.~\ref{fig:JSP_Tradeoffs}B) might be obtained by tuning the network at coexistence between localized and delocalized phases.
%The system is expected to be in the signaling states when the ligand is bound to the receptor and 
%with signaling and non-signaling states.  
\label{fig:BP}}
\end{figure}

\subsection{ Thermodynamics of concentration measurements} 

Exploring the tradeoffs between entropy consumption and the ability of a biochemical apparatus to sense concentrations has been a topic of recent interest~\cite{Mehta:2012ji,Lang:2014ir,Govern:2014ef,Barato:2014fg,Barato:2013ew}. 
A recent paper by Lang et al~\cite{Lang:2014ir} puts forward a general phenomenological model (Fig.~\ref{fig:BP}) to understand the tradeoffs between concentration sensing and entropy consumption. 
The model system is composed of a non signaling chemical state coupled to an array of signaling states. The rates for transition between signaling and non signaling states depends on the concentration of the ligand. The transition from the non signaling state to the signaling states is driven by  ligand binding events. 
For constant rates $k_{cw}$ and $k_{ccw}$ for transitions between the signaling states, the network in Fig~\ref{fig:BP} (and its generalizations with additional signaling states) qualitatively supports the same dynamical 
phases described in the previous sections~\cite{Vaikuntanathan2014}. The transition rates between the non-signaling and the signaling links play the role of the heterogeneity or the $h$ link. 

The concentration of the ligand can be inferred from measurements of the time spent by the system in the signaling states. 
The uncertainty in the concentration measured has been worked out to be~\cite{Lang:2014ir} 
\begin{equation}
\label{eq:MehtaMLE}
\frac{\langle (\delta c)^2\rangle }{\bar c^2}=\frac{1}{\bar N}\left[1+\frac{\langle (\delta \tau_s)^2\rangle}{\bar \tau^2_s} \right ]
\end{equation}
where $\bar N$ is the number of ligand binding events, $\bar \tau_s$ is the mean time spent in the signaling state after binding a ligand and $\langle (\delta \tau_s)^2\rangle$ is the variance of the time spent in the signaling states.

Can the variance of the time spent in the signaling states, and hence the uncertainty, be reduced at the cost of energy consumption? Lang et al~\cite{Lang:2014ir} find this to be case. In fact, they find a trade-off curve resembling Fig.~\ref{fig:JSP_Tradeoffs}B where a large reduction in uncertainty can obtained at a small energy cost followed by a regime of small further reductions in uncertainty at a large energy cost. This analogy suggests that tradeoffs similar to those discussed here for kinetic proofreading and search strategies might be accessible to signaling networks. A signaling network at the cusp of dynamic phase transition can remain localized near the non signaling states in the absence of any ligand binding events. A minor perturbation, such as a ligand binding event, can induce the system to undergo a transition to the delocalized signaling state. 
The trade-off between reliability of the concentration readout and entropy consumption might be optimal closer to such a coexistence point. 

Additional support for the unifying theme of dynamical phases comes from recent work on variable sensitivity and programmability of concentration sensing \cite{Hartich:2015vp,Mehta:2012ji}. \cite{Hartich:2015vp} found that the sensitivity to concentrations can be modulated, closely paralleling variable proofreading regimes found in \cite{Murugan:2014uq}. The parallels extend to `anti-proofreading’ regimes of \cite{Murugan:2014uq} where the response to changing binding energies (or ligand binding in \cite{Hartich:2015vp}) is reduced as compared to equilibrium. In \cite{Murugan:2014uq}, some of these `anti-proofreading’ regimes were interpreted in terms of modular sub-networks which can in turn be interpreted as designer phase coexistence. The general relationship between modular design of non-equilibrium mechanisms and dynamical phases deserves further study.

\subsection{Possible implications for ultra sensitivity} 

Non-equilibrium driving forces also play important roles in sensory adaptation~\cite{Lan:2012in} and ultra sensitivity~\cite{Tu:2008vn}. The framework of localized and de-localized phases might indeed apply to these other processes. The localized phases can  be a generic route used by biochemical systems to reinforce information encoded in the equilibrium landscape.

Consider for instance ultra sensitive switches \cite{Tu:2008vn,Lin:2014jf} found in E. coli chemotaxis and other biochemical signaling motifs. Existing modeling approaches rely on an equilibrium allosteric framework. However, recent experimental \cite{Korobkova:2006vw,Lin:2014jf} and theoretical results~\cite{Tu:2008vn} suggest that non-equilibrium forces might play a crucial role in ultra sensitive switches 
The landscape imposed by the equilibrium allosteric interactions can be reinforced with non equilibrium fluxes in a localized phase.  
Further an ultra sensitive switch positioned at the cusp of a dynamic phase transition can fruitfully utilize non equilibrium driving forces to rapidly go between different dynamical regimes. Indeed, the enhanced fluctuations in the ultra sensitive E. Coli flagella motor as it switches from the clockwise rotational mode to the counter clockwise rotation mode is reminiscent of a delocalized state. It will be interesting to study well characterized ultra sensitive systems and investigate whether dynamical phase transitions serve as an important organizing principle. 

\section{Acknowledgments}

We gratefully acknowledge useful discussions with Michael Brenner, Aaron Dinner, Todd Gingrich, David Huse,  Stan Leibler, Pankaj Mehta, Matthew Pinson, Luca Peliti, Mike Rust, Mikhail Tikhonov and Thomas Witten. SV acknowledges funding from the University of Chicago.

%\subsection{Variable sensitivity / Modular design principle}

%\bibliographystyle{unsrt}
%\bibliography{Papers3_Bibtex.bib,ReferencesSuri1.bib,References_Suri.bib}

% from the bbl file

\end{document}